\documentclass[conference]{IEEEtran}
\IEEEoverridecommandlockouts
\usepackage{cite}
\usepackage{amsmath,amssymb,amsfonts}
\usepackage{graphicx}
\usepackage{textcomp}
\usepackage{url}
\usepackage{xcolor}
\usepackage{tightenum}
\usepackage{placeins}
\usepackage{booktabs}
\usepackage{color}
\usepackage{xcolor}
\usepackage{subcaption}
\usepackage{graphics,graphicx,color}
\usepackage{multirow}
\usepackage{xspace}
\usepackage{algpseudocode}
\usepackage[ruled,vlined,linesnumbered]{algorithm2e}
\usepackage{enumitem}
\usepackage{ulem}
\usepackage{setspace}
\SetVlineSkip{0.05em}
\AtBeginEnvironment{figure}{\setlength{\belowcaptionskip}{-1em}\centering}
\AtBeginEnvironment{subfigure}{\centering}
\AtBeginEnvironment{figure*}{\setlength{\belowcaptionskip}{-1em}\centering}
\usepackage[compact]{titlesec}

\definecolor{teal}{rgb}{0.0, 0.6, 0.004}
\definecolor{brown}{rgb}{0.804, 0.0, 0.0}
\usepackage{tcolorbox}
\usepackage{ulem}

\newcommand{\xxx}{Themis\xspace}
\newcommand{\coverage}{Sensitivity Convergence Coverage\xspace}
\newcommand{\fuzzer}{Sensitivity Maximizing Fuzzer\xspace}

\newcommand{\calculator}{Sensitivity Calculator\xspace}
\newcommand{\sampler}{Sensitivity Sampler\xspace}

\newcommand{\fdc}{fault detection capability\xspace}

\newcommand{\lowcorr}{0.70\xspace}
\newcommand{\highcorr}{0.95\xspace}
\newcommand{\lowcorrrelated}{-0.89\xspace}
\newcommand{\highcorrrelated}{0.59\xspace}
\newcommand{\numfault}{3.78X\xspace}

\newcommand{\lowaccxxx}{0.21\%\xspace}
\newcommand{\highaccxxx}{8.77\%\xspace}
\newcommand{\lowaccbaselines}{0.01\%\xspace}
\newcommand{\highaccbaselines}{3.48\%\xspace}
\newcommand{\avgaccx}{14.7X\xspace}

\newcommand{\vggacc}{11.56\%\xspace}

\newcommand{\note}{black}
\newcommand{\curr}{black}

\def\BibTeX{{\rm B\kern-.05em{\sc i\kern-.025em b}\kern-.08em
    T\kern-.1667em\lower.7ex\hbox{E}\kern-.125emX}}
\begin{document}

\title{Themis: Automatic and Efficient Deep Learning System Testing with Strong Fault Detection Capability}

\author{
\IEEEauthorblockN{Dong Huang\textsuperscript{}}
\IEEEauthorblockA{University of Hong Kong\\
dhuang@cs.hku.hk}
\and
\IEEEauthorblockN{Tsz On Li\textsuperscript{}}
\IEEEauthorblockA{University of Hong Kong\\
toli2@cs.hku.hk}
\and
\IEEEauthorblockN{Xiaofei Xie}
\IEEEauthorblockA{Singapore Management University\\
xiaofei.xfxie@gmail.com}
\and
\IEEEauthorblockN{Heming Cui}
\IEEEauthorblockA{University of Hong Kong\\
heming@cs.hku.hk}
}

\maketitle

\begin{abstract}
  Deep Learning Systems (DLSs) have been widely applied in safety-critical tasks such as autopilot.  However, when a perturbed input is fed into a DLS for inference, the DLS often has incorrect outputs (i.e., faults). DLS testing techniques (e.g., DeepXplore) detect such faults by generating perturbed inputs to explore data flows that induce faults. Since a DLS often has infinitely many data flows, existing techniques require developers to manually specify a set of activation values in a DLS's neurons for exploring fault-inducing data flows. Unfortunately, recent studies show that such manual effort is tedious and can detect only a tiny proportion of fault-inducing data flows.

In this paper, we present \xxx, the first  automatic DLS testing system, which attains strong fault detection capability  by ensuring a full coverage of fault-inducing data flows at a high probability.  \xxx carries a new workflow for automatically and systematically revealing data flows whose internal neurons' outputs vary substantially when the inputs are slightly perturbed, as these data flows are likely fault-inducing. We evaluated \xxx on ten different DLSs and found that on average the number of faults detected by \xxx was \numfault more than four notable DLS testing techniques. By retraining all evaluated DLSs with the detected faults, \xxx also increased (regained) these DLSs' accuracies on average \avgaccx higher than all baselines.

\end{abstract}


\section{Introduction}\label{sec:intro}

Deep Learning Systems (DLSs) have been widely 
applied in safety-critical tasks such 
as autopilot and 
smart cities~\cite{NEURIPS2018_69386f6b,
pei2017deepxplore,
ma2018deepGauge,
gerasimou2020importance,
kim2019guiding}. However, when a DLS (e.g., an autopilot system) is deployed in  a real-world environment (e.g., a crowded city), the 
DLS's input (e.g., a road condition 
image of the crowded city) is often perturbed by 
environmental noise such as 
raindrops and fog effects, causing incorrect values and disaster
~\cite{pei2017deepxplore,ma2018deepGauge,
gerasimou2020importance,kim2019guiding,huang2023neuronsensitivityguidedtest,huang2024adversarial,huang2023featuremaptestingdeep}. 
The incorrect output value of a DLS caused by 
perturbation on the DLS's input is defined as the DLS's 
fault~\cite{harel2020neuron}.

Since a DLS's faults greatly 
undermine the DLS's reliability, 
a DLS must be systematically 
tested in order to detect as many faults as 
possible 
~\cite{pei2017deepxplore,
ma2018deepGauge,
gerasimou2020importance,
kim2019guiding}.  
These faults are essential 
for further developing a DLS (i.e., improving the 
DLS's accuracy by retraining the DLS 
with perturbed inputs which lead to faults). The rationale of existing 
DLS testing is 
inspired by conventional 
software testing. In conventional 
software testing~\cite{sutton2007fuzzing,
manes2019art}, a testing technique generates a 
set of inputs (known as the \textit{test set}) to  
explore diverse 
data flows of a software (e.g., statements or branches), 
especially the data flows which  
likely lead to the software's faults such as
exceptions or crashes (these data flows are 
called ``fault-inducing data flows''). 
When all  
data flows of the software have  
been explored (i.e.,  the 
test coverage metric adopted by 
the testing technique, such as code coverage, reaches 
100\%), the testing process is considered completed, and the testing technique stops.

Similarly, existing DLS testing techniques (e.g., DeepXplore~\cite{pei2017deepxplore}, 
DeepGauge~\cite{ma2018deepGauge},
DeepImportance~\cite{gerasimou2020importance}, and Surprise Adequacy for 
Deep Learning System~\cite{kim2019guiding}) generate  
a test set (a set of perturbed inputs) to explore a DLS's 
data flows, where  
the DLS's data flow is defined as a 
set of numerical values: each of the 
numerical value is one of  
the DLS's neurons' output value (a DLS's 
neuron is a ``function'', and the output value of a neuron 
is called ``activation 
value''~\cite{pei2017deepxplore,ma2018deepGauge}) 
corresponding to a DLS's input. A DLS testing instance generates noise test set 
by perturbing test set with the same type of environmental 
noise (e.g., raindrop effect) of various levels of
noise strength (e.g., various raindrop 
densities).

However, since the activation values of a DLS's neurons    
are discrete numbers from negative infinity 
to positive infinity, generating a test set to     
make a DLS's neurons output all possible sets of 
activation values 
is prohibitively inefficient. To mitigate 
such inefficiency in DLS testing, for each type of environmental effects (noise), 
existing techniques 
(DeepXplore~\cite{pei2017deepxplore}, 
DeepGauge~\cite{ma2018deepGauge},
DeepImportance~\cite{gerasimou2020importance},
and Surprise Adequacy for 
Deep Learning System~\cite{kim2019guiding}) 
require a DLS's developer to manually specify likely
sets of fault-inducing activation values in all neurons.

{\color{\curr} For example, DeepXplore~\cite{pei2017deepxplore} requires developers to manually partition a neuron’s outputs into two segments (i.e., outputs larger than a threshold and outputs smaller than a threshold, where the threshold has to be manually specified by the developers). Then, DeepXplore ~\cite{pei2017deepxplore} generates test inputs to cover each of the two segments for all neurons. Nevertheless, recent work ~\cite{harel2020neuron} showed that manually specifying the threshold is error-prone, and such manual effort is often overly coarse-grained, identifying a DLS's all activation values that may induce faults for each type of environmental effects is fundamentally difficult, and not all values will induce faults on all neurons. Worse, existing 
DLS testing techniques~\cite{pei2017deepxplore,gerasimou2020importance} often can't attain strong fault detection capability (strong fault detection 
capability means the correlation 
between the error rate of a DLS
and the number of the DLS's faults 
detected is larger than 0.7). }

We believe that the root cause behind the limitations of the existing techniques is 
that the triggering condition 
and the total number of a DLS's fault-inducing 
data flows are unknown~\cite{pei2017deepxplore,
ma2018deepGauge,harel2020neuron,li2019structural,
kim2019guiding,gerasimou2020importance}. Hence, existing techniques 
inevitably require manual expert effort in inferring 
a set of fault-inducing data flows, in order to 
guide the exploration of a DLS's data flows 
and the computation of 
test coverage (i.e., the proportion of the 
fault-inducing data flows being explored from the DLS 
during testing). 
Unfortunately, despite these advancements~\cite{pei2017deepxplore,
ma2018deepGauge,harel2020neuron,li2019structural,
kim2019guiding,gerasimou2020importance}, in principle, manual effort is far away from   
guaranteeing that these techniques can 
explore a full coverage of fault-inducing 
data flows from a DLS, 
while the full coverage is an essential and sufficient condition for a testing 
technique to attain strong fault-detection 
capability~\cite{harel2020neuron,li2019structural}. 
Overall, an automated  
DLS testing technique that can theoretically explore  
a full coverage of fault-inducing data flows  
is highly desirable but vacant.

\begin{figure}[]
    \centering
    \includegraphics[width=1\columnwidth]{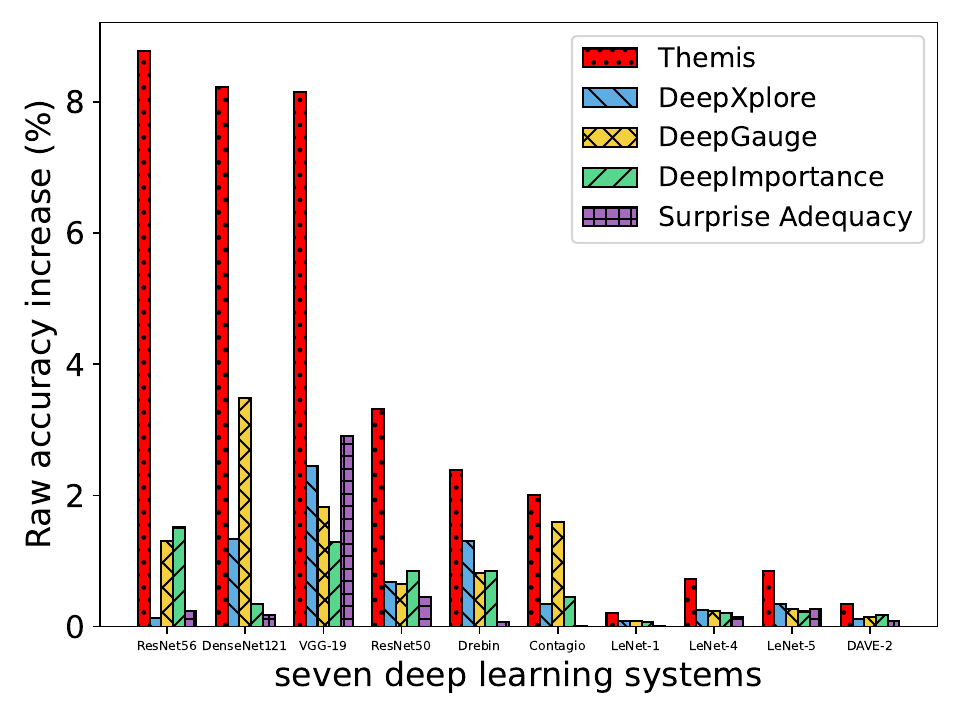}
    \caption{
    Increase in 
    a DLS's accuracy by 
    retraining the DLS with faults 
    detected by \xxx and the baselines.
    The increases in DLS's accuracy brought by 
    \xxx were on average \avgaccx more than 
    the increases in DLS's accuracy brought by baselines.} 
    \label{fig:intro}
\end{figure}

The main observation of the paper is that, a fault-inducing data flow in a DLS is often  
\textit{sensitive} to perturbation on a DLS's input.
We denote the 
data flow with respect to a DLS $M$ and an input $I$ as $DF(M,I)$. 
For a DLS ($M$), a clean input ($I$) and its perturbed input ($I'$), 
we consider 
a data flow $DF(M,I')$ \textit{sensitive} 
to perturbation if the difference of $DF(M,I')$ and $DF(M,I)$
(denoted as $DF(M,I')-DF(M,I)$) is large,  
given that $I'$ and $I$ just have slight 
difference (e.g., $I'$ and $I$ differ by light 
raindrops). 
Intuitively,  
sensitive $DF(M,I')$ inevitably causes  
$M(I')$ and $M(I)$ to be different, confirmed in our evaluation (Figure~\ref{fig:efficiency}b).

With this observation, we 
present \xxx, a DLS testing technique that 
systematically explores fault-inducing data flows 
guided by data flows' sensitivity: the difference between  
$DF(M,I')$ and $DF(M,I)$ with respect to the difference between 
$I'$ and $I$. 
\xxx leverages math optimization techniques 
(e.g., Gradient Descent) to adjust the intensity of 
the perturbation (e.g., raindrops' densities) 
on $I$, in order to generate a new test set of  
$I'$, which maximizes $DF(M,I')-DF(M,I)$. 
Hence, \xxx maximizes the 
likelihood that $DF(M,I')$ will lead to a fault. 
With this new workflow, \xxx 
explores faults from a DLS without the necessity of either manual effort or 
exploring all data flows of a DLS. One major challenge for \xxx to  
achieve strong fault detection 
capability is how \xxx infers the coverage of 
fault-inducing data flows being explored 
by \xxx's generated test set during testing 
(i.e., how \xxx computes the test coverage metric).  
This is because the actual 
number of a DLS's fault-inducing data flows on all inputs 
is unknown. 

{\color{\curr}To tackle this challenge, 
we summarize existing AI
theories~\cite{NEURIPS2018_69386f6b,
long2018pde,
robert1998discretization,
zheng2016improving} to 
show that, a DLS's $DF(M,I')-DF(M,I)$ on all inputs often follows normal 
distribution: for all clean inputs, most noise added to these 
inputs will lead to similar influence on $M$'s final outputs, and only a tiny portion of noise will lead to outlying influence.
Hence, once \xxx's workflow infers that $DF(M,I')-DF(M,I)$  
converges to a normal distribution driven by \xxx's generated 
test set, according to these theories, \xxx has explored a statistically 
full coverage of fault-inducing data flows (i.e., only a tiny portion of fault-inducing noise and their data flows were missed by \xxx), and \xxx's testing instance 
can complete.  
Our paper derives a proof (\S\ref{sec:analysis}) to show that \xxx's workflow can achieve statistically full coverage (i.e., achieving the ``fault detection capability with high probability'' in this paper). }

We implemented \xxx on Mindspore and 
compared \xxx against four recent and notable DLS testing techniques  
(Deepexplore~\cite{pei2017deepxplore}, 
DeepGauge\cite{ma2018deepGauge}, 
DeepImportance~\cite{gerasimou2020importance} and 
Surprise Adequacy for Deep 
Learning System~\cite{kim2019guiding}), which have been deployed in the Mindspore security framework and have been evaluated by several AI developers. We evaluated \xxx and these baselines on ten popular Deep Learning models 
(e.g., LeNet~\cite{lecun1998gradient}, 
VGG~\cite{simonyan2014very}, ResNet~\cite{he2016deep}) trained on six public datasets (Cifar10~\cite{cifar10}, ImageNet~\cite{deng2009imagenet}, Driving~\cite{udacity}, Contagio/VirusTotal~\cite{contagio}, Drebin~\cite{arp2014drebin} and MNIST~\cite{lecun1990handwritten}), which cover a complete set 
of datasets evaluated by the baselines~\cite{pei2017deepxplore, 
ma2018deepGauge, 
gerasimou2020importance, 
kim2019guiding}.
Evaluation shows that:

\begin{itemize}[leftmargin=1em]
    \item {\xxx consistently achieved strong \fdc. \xxx achieved a high correlation~\cite{harel2020neuron} 
    (\lowcorr to \highcorr) for all the DL models, while the 
    baselines' correlation varied from \lowcorrrelated to 
    \highcorrrelated (Table~\ref{tab:corr}).}

    \item {
    \xxx detected \numfault more 
    faults than the baselines, 
    when \xxx's and the baselines' test coverage reach 100\% 
    .}

    \item {By retraining the DL models with faults detected in 
    testing, \xxx increased (regained) the DL models' accuracy by 
    \lowaccxxx to \highaccxxx, while baselines 
    increased the DL models' accuracy by \lowaccbaselines to 
    \highaccbaselines. Overall, \xxx increased the DL models' accuracy on 
    average \avgaccx higher than 
    the baselines (Figure~\ref{fig:intro}).}
   
    \item {\xxx has been integrated into the Mindspore security framework, enhancing the testing of hundreds to thousands of DNNs in real-world scenarios.}

\end{itemize} 

Our main contribution is \xxx, the first 
automatic DLS testing technique  
which can empirically attain strong fault 
detection capability for perturbed inputs. 
The key novelty is \xxx's new workflow for systematically 
exploring a DLS's  
fault-inducing data flows and computing its  
test coverage metric without manual effort, by leveraging our observation 
that most fault-inducing 
data flows are sensitive. Our theoretical 
analysis (\S\ref{sec:analysis}) 
shows that \xxx can explore a full coverage 
of fault-inducing data flows at high probability (95\%), 
so \xxx empirically attained strong fault detection 
capability for all the evaluated 
DLSs (Table~\ref{tab:corr}). 

\section{Background and Related Work}\label{sec:back}

\subsection{Deep Learning Testing}
DeepXplore~\cite{pei2017deepxplore} proposes the first white-box coverage criteria, i.e., Neuron Coverage, which calculates the percentage of activated neurons. A differential testing approach is proposed to detect the errors by increasing NC. DeepGauge~\cite{Ma2018DeepGaugeMT} then extends NC and proposes a set of more fine-grained coverage criteria by considering the distribution of neuron outputs from training data. Inspired by the coverage criteria in traditional software testing, some coverage metrics~\cite{Sun2018ConcolicTF, Ma2019DeepCTTC, Hu2019DeepMutationAM, Ma2018DeepMutationMT, gerasimou2020importance, Tian2017DeepTestAT, Xie2019DeepHunterAC} are proposed. DeepCover~\cite{Sun2018ConcolicTF} proposes the MC/DC coverage of DNNs based on the dependence between neurons in adjacent layers. DeepCT~\cite{Ma2019DeepCTTC} adopts the combinatorial testing idea and proposes a coverage metric that considers the combination of different neurons at each layer. DeepMutation~\cite{Ma2018DeepMutationMT} adopts the mutation testing into DL testing and proposes a set of operators to generate mutants of the DNN. Furthermore, Sekhon et al.\cite{Sekhon2019TowardsIT} analyzed the limitation of existing coverage criteria and proposed a more fine-grained coverage metric that considers both the relationships between two adjacent layers and the combinations of values of neurons at each layer. Based on the neuron coverage, DeepPath ~\cite{Wang2019DeepPathPT} initially proposes the path-driven coverage criteria, which considers the sequentially linked connections of the DNN. IDC~\cite{gerasimou2020importance} adopts the interpretation technique to select the important neurons in one layer. Based on the training data, it then groups the activation values of important neurons into a set of clusters and uses the clusters to measure the coverage. Kim et al.\cite{Kim2018GuidingDL} proposed the coverage criteria that measure the surprise of the inputs. The assumption is that surprising inputs introduce more diverse data such that more behaviors could be tested. Surprise metric measures the surprise score by considering all neuron outputs of one or several layers. It is still unclear how the surprise coverage (calculated from some layers) is related to the decision logic. Xie et al.\cite{Xie2022NPCNP} propose Neuron Path Coverage, a novel, and interpretable coverage criterion aimed at characterizing the decision logic of models.

\subsection{Existing DLS testing are unautomated}\label{sec:back:testing}


DLS testing techniques (e.g., Deepexplore~\cite{pei2017deepxplore}, 
DeepGauge\cite{ma2018deepGauge}, 
DeepImportance~\cite{gerasimou2020importance}, and 
Surprise Adequacy for Deep 
Learning System~\cite{kim2019guiding}) are proposed 
to detect a DLS's faults. A testing instance includes 
a pretrained DLS $M$ (e.g., 
an autopilot system), an arbitrary input $I$ fed to  
$M$ for inference 
(e.g., a road condition image), and the same  
type of environmental 
noise with arbitrary noise 
strength $E(\theta)$  (e.g., raindrop effect $E$,  
with arbitrary raindrop 
size $\theta$, where $\theta$ is within a given range such 
as the real-world raindrop's sizes) 
which may be present 
on $I$. For any $M(I+E(\theta))$ which 
is different from $M(I)$, the $M(I+E(\theta))$ is considered 
a DSL's fault. 

Note that an incorrect $M(I+E(\theta))$ is considered as a 
fault instead of a failure because existing DLS testing techniques 
aim to test deep learning models  
of a DLS rather than the entire 
DLS (i.e., both deep learning models and the software code 
of a DLS). Since in existing DLSs (e.g., industrial 
autonomous systems such as Autoware~\cite{kato2018autoware} 
and Apollo~\cite{apollo}), 
deep learning models are intermediate components, the 
incorrect output of these deep learning models are 
regarded as the ``fault'' of a  DLS~\cite{harel2020neuron}. 

These techniques generate perturbed inputs 
to make a DLS's neurons output diverse activation values, 
in order to explore the DLS's data flows 
(\S\ref{sec:intro}).
Specifically, denote 
$N_{i},i=1,..,n$ as the $i$th neuron of a DLS, and 
$N_{i}(I+E(\theta)) \in \mathbb{R}$ 
as the activation values of the $i$th neuron 
corresponding to input $I+E(\theta)$.
A DLS's data flow is defined as a set of 
numerical values, where each numerical 
value is each neuron's activation 
of the DLS (i.e., a data flow is defined as 
$\{N_{i}(I+E(\theta))\}$). Existing work 
generates diverse $I+E(\theta)$, 
in order to trigger diverse 
data flows.

Since activation values 
are discrete numbers from negative infinity to positive 
infinity, existing techniques proposed various 
heuristic rules for a DLS's developers 
to manually specify fault-inducing data 
flows. For instance, 
DeepXplore~\cite{pei2017deepxplore} requires 
a DLS's developer to divide 
the activation values of neurons 
into two segments (e.g., values larger than zero 
and values smaller than zero), such that DeepXplore only 
needs to generate at minimal two perturbed inputs 
(we denote these inputs as $I_{1}+E_{1}(\theta_{1})$ 
and $I_{2}+E_{2}(\theta_{2})$), 
to cover the segments (e.g., 
$N_{i}(I_{1}+E_{1}(\theta_{1})) \geq 0\ \forall i$ and 
$N_{i}(I_{2}+E_{2}(\theta_{2})) > 0\ \forall i$). 
Other DLS testing techniques also proposed 
similar heuristic rules for developers to 
specify fault-inducing data flows. 
DeepGauge~\cite{ma2018deepGauge} requires the user to 
divide activation values into several segments, where 
the segments that contain activation 
values towards zero have smaller intervals (because most 
inputs of a DLS make the neurons' output activation values 
close to zero).
DeepImportance~\cite{gerasimou2020importance} 
requires a DLS developer to specify 
fault-inducing activation values for ``important'' neurons 
only (the ``important'' neurons are identified 
via a popular technique in DL called 
``Layer-wise Relevance Propagation''~\cite{montavon2019layer}). 
Surprise Adequacy for Deep 
Learning System~\cite{kim2019guiding} 
requires a DLS's developer to specify fault-inducing 
activation values for neurons in softmax layers only.

\subsection{Fault-inducing data flows have to be 
automatically identified}\label{sec:back:prob}

Overall, 
these techniques allow  
a DLS to be tested within a reasonable time (e.g., several 
minutes). However, substantial 
studies~\cite{harel2020neuron,li2019structural} showed that 
manual effort in specifying fault-inducing 
data flows often causes DLS testing 
to have unsatisfactory 
fault detection capability because 
fault-inducing data flows are often unknown 
(see \S\ref{sec:intro}). To automatically identify fault-inducing data flows, 
our observation is that 
for any $I$ and $E(\theta)$, 
the $N_{i}(I+E(\theta)), 1 \leq i \leq n$ which 
induces a fault usually 
has a large difference with $N_{i}(I), 
1 \leq i \leq n$ (i.e., the value of 
$sum_{i=1}^{n} |N_{i}(I+E(\theta)) - N_{i}(I)|$ 
is large, we call this value as the \textit{sensitivity} of a DLS's 
data flow on perturbed inputs). It is because 
large sensitivity often results in a large difference between  
$M(I+E(\theta))$ and $M(I)$ (i.e., $M(I+E(\theta))$ 
is likely to be a fault). This observation is also confirmed 
in our evaluation (\S\ref{sec:eval:fdc}).  
Based on 
this observation, we propose an input generation technique 
guided by \textit{sensitivity} (\fuzzer 
in \S\ref{sec:overview}).


\subsection{Computing test coverage automatically}\label{sec:back:mcmc} 

\begin{figure*}
    \centering
    \includegraphics[width=0.9\textwidth]{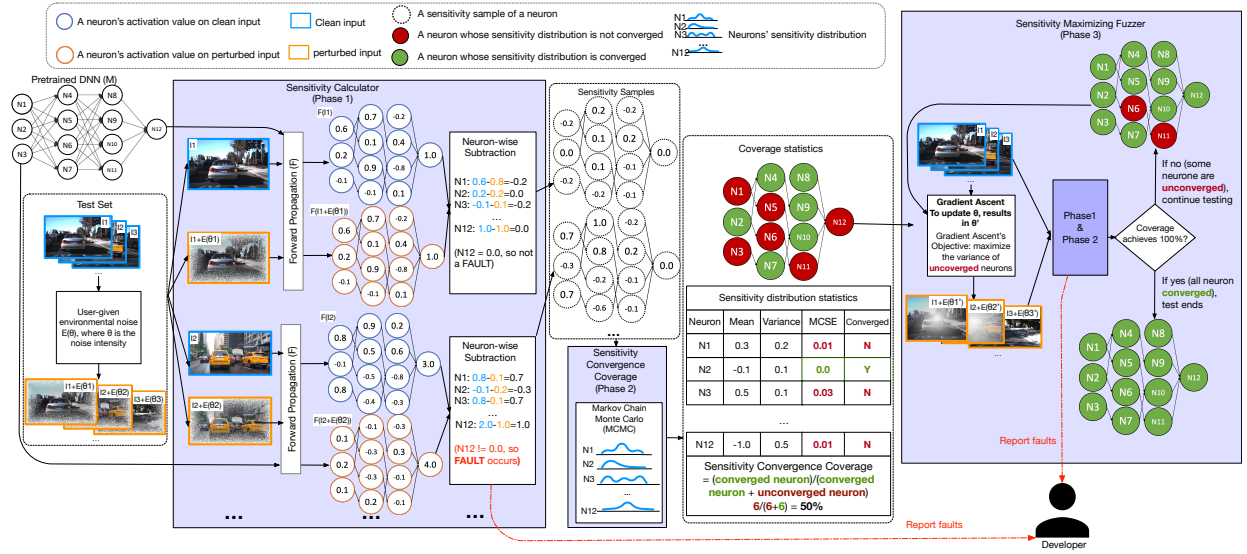}
    \caption{\xxx's architecture. \xxx's components 
    are shaded in purple. The DNN's neurons 
    (e.g., N2) are colored in 
    green if their MCSE value 
    (as shown in the ``Sensitivity 
    distribution statistics'') 
    equals 0.0.}
    \label{fig:arch}
\end{figure*}

{\color{black}
Test coverage metric measures the proportion 
of fault-inducing data flows  
explored. Ideally, when test coverage metric 
reaches 100\%, all fault-inducing data flows 
of a DLS are explored~\cite{harel2020neuron,li2019structural}. However, since the actual number of 
fault-inducing data flows in a DLS is 
unknown, proposing 
an accurate test coverage metric 
is an open challenge~\cite{harel2020neuron,li2019structural}. 
Existing testing techniques~\cite{NEURIPS2018_69386f6b,
pei2017deepxplore,
ma2018deepGauge,
gerasimou2020importance,
kim2019guiding}  
mitigate this challenge by requiring  
developers to specify a set of fault-inducing 
data flows. However, recent studies 
show that such manual effort is tedious and error-prone 
(\S\ref{sec:back:testing}). 

To solve this challenge, \xxx leverages the sensitivity of  
a DLS's data flow to infer the 
test coverage. We summarize the existing theories 
to show that for any DLS, $N_{i}(I+E(\theta))-N_{i}(I)$ follows a 
normal distribution. 
First, existing 
theories show that DNN is an ordinary differential 
equation~\cite{NEURIPS2018_69386f6b,
long2018pde}, and DNN's neurons are differential 
operators of an ordinary differential equation. 
Second, when random 
noise is present on an ordinary differential 
equation's input, the variation on the output values 
of the ordinary differential equation's differential 
operators often follows a normal 
distribution~\cite{robert1998discretization,
zheng2016improving}. By combining the first and the second theories, we can derive that when random noise is present on a DLS's input, the variation of the DLS's activation values (i.e., $N_{i}(I+E(\theta)) - N_{i}(I)$ 
defined in \S\ref{sec:back:testing}) often follows a normal distribution (we denote $ND_{i}$ as the normal distribution, and 
we call $\hat{ND_{i}}$ as ``sensitivity distribution'' in Figure~\ref{fig:arch}). 
We also confirmed these observations in our evaluation (Figure~\ref{fig:mcmcmean}b). 

Since $\hat{ND_{i}}$ consists of the frequency of all $N_{i}(I+E(\theta)) - N_{i}(I)$ values including the fault-inducing data flows $\{N_{i}(I+E(\theta))\}$, the condition that $\hat{ND_{i}}$ being identified as $ND_{i}$ implies all fault-inducing data flows are also identified (we carry a more detailed analysis in \S\ref{sec:analysis}). Based on this observation, we proposed a test coverage metric based on the distribution of $ND_{i}$  (\coverage in \S\ref{sec:overview}).}

\section{Overview}\label{sec:overview}


This section presents the 
architecture and workflow of 
\xxx. Figure~\ref{fig:arch} shows \xxx's major 
components (namely \calculator, \coverage and 
\fuzzer) and these components' workflow. 
To ease discussion, this section 
uses the notations defined 
in \S\ref{sec:back:testing}: 
given a DLS ($M$), 
a set of inputs ($I$) for the DLS to perform inference and
environmental 
noise to be applied on $I$ ($E(\theta)$), 
\xxx aims to detect the $M(I+E(\theta))$ which 
is not equal to $M(I)$ (such an $M(I+E(\theta))$ is a fault).

To do so, \xxx first perturbs each input of $I$ 
with environmental noise $E(\theta)$ (for each input in $I$, 
$\theta$ is a randomly chosen value within a given 
range). Then, \xxx feeds both 
$I$ and $I+E(\theta)$ to \textbf{\calculator (Phase 1)}, to compute 
the difference between the 
outputs of each $M$'s neurons 
with respect to these inputs 
(i.e., $N_{i}(I+E(\theta))-N_{i}(I)$ defined 
in \S\ref{sec:back:testing},  
which are the $i$th neuron's outputs with respect to 
$I$ and $I+E(\theta)$ respectively). 
\calculator also computes $M(I)$ and $M(I+E(\theta))$, and 
reports faults to DLS developers (i.e., $M(I+E(\theta))$ which 
is not equal to $M(I)$). 

{\color{\note}
Note that \xxx needs to compute 
both $N_{i}(I)$ and $N_{i}(I+E(\theta))$, 
while existing DLS testing 
techniques only compute $N_{i}(I+E(\theta))$, 
even though all these techniques aim to trigger 
fault-inducing $N_{i}(I)$. 
It is because \xxx leverages the theoretical implication behind   
$N_{i}(I+E(\theta))-N_{i}(I)$ to guide the detection 
of fault-inducing $N_{i}(I+E(\theta))$, in 
order to automatically identify fault-inducing 
data flows (see \S\ref{sec:back:prob}).  
On the other hand, existing techniques 
require fault-inducing $N_{i}(I+E(\theta))$ to be 
specified by DLS developers, which has been proven 
tedious and error-prone (\S\ref{sec:back:mcmc}).
}

\calculator then passes 
$N_{i}(I+E(\theta))-N_{i}(I)$ values 
to \textbf{\coverage (Phase 2)}, which infers a 
distribution from the values (i.e., $\hat{ND_{i}}$ defined 
in \S\ref{sec:back:mcmc}) for each neuron (i.e., each $i \in n$). 
\coverage then determines whether $\hat{ND_{i}}$ 
converges to a normal 
distribution and returns a test coverage value accordingly. 
Specifically, for each neuron in a DLS, 
\coverage feeds the $N_{i}(I+E(\theta))-N_{i}(I)$ values 
to Monte Carlo 
Markov Chain~\cite{geyer1992practical} 
(MCMC, 
a popular statistical technique for inferring 
probability distributions from a set of numerical values). 

MCMC then outputs both $\hat{ND_{i}}$ (the 
distribution inferred by MCMC) and the  
Monte Carlo Standard Error 
(MCSE, a metric to measure the potential 
error between the inferred distribution and 
the ground-truth normal distribution) value corresponding to 
$\hat{ND_{i}}$. \coverage thus  
determines whether $\hat{ND_{i}}$ converges to a normal 
distribution based on the MCSE value: 
If MCSE equals zero within a confidence interval 
$\alpha$ (\xxx's default value of 
$\alpha$ is 95\%, a standard value 
of setting a confidence 
interval~\cite{hosmer1992confidence,altman2011obtain}), 
then \coverage considers $\hat{ND_{i}}$ 
converges to a normal distribution at a
high probability (i.e., 95\%). Otherwise, 
$\hat{ND_{i}}$ is considered not converged.

\coverage then computes the test coverage metric 
as the proportion of neurons whose $\hat{ND_{i}}$ 
converges. For instance, in Figure~\ref{fig:arch}, 
after \coverage computes $\hat{ND_{i}}$ for 
the twelve neurons, six 
neurons among the twelve neurons 
(i.e.,N2, N4, N7, N8, N9 and 
N10 which are colored in green) have MCSE as 0.0 
(i.e., their $\hat{ND_{i}}$ are converged), 
while the rest 
of the six neurons of the DLS have MCSE large than 0.0 
(i.e., their $\hat{ND_{i}}$ are not converged). 
Hence, \coverage computes the test coverage 
as 50\%.

{\color{\note}
\coverage addresses an open challenge 
of DLS testing: accurately inferring the 
proportion of faults undetected from a DLS 
(i.e., computing a test coverage metric), even the 
total number of the DLS's faults is unknown. 
Specifically, \coverage does so by 
inspecting $\hat{ND_{i}}$'s convergence condition, which is 
correlated with the proportion of the DLS's faults being 
detected (\S\ref{sec:back:mcmc}). Our theoretical 
analysis showed that when \coverage reaches 
100\%, all fault-inducing data flows are identified 
(i.e., explored) at high probability (95\%). In contrast, 
existing works mitigate this challenge by requiring 
DLS developers to specify a set of fault inducing data 
flows, such that the test coverage metric is computed as 
the specified fault inducing data 
flows being explored. } d

After that, \coverage passes the neurons whose  
$\hat{ND_{i}}$ is converged to \xxx's \textbf{\fuzzer (Phase 3)}. 
\fuzzer then leverages math optimization 
techniques (e.g., Gradient 
Ascent)  
to generate a new test set,  
to explore unexplored fault-inducing 
$N_{i}(I+E(\theta))$. Specifically, 
\fuzzer iteratively 
adjusts $\theta$ of $E$ (we denote the adjusted 
$\theta$ as $\theta'$), such that $I+E(\theta')$ 
maximizes the sum 
of $N_{i}(I+E(\theta))-N_{i}(I))$ for all neurons 
whose $\hat{ND_{i}}$ are not converged (i.e., 
N2, N4, N7, N8, N9 and N10 in Figure~\ref{fig:arch}). \fuzzer 
does so to ensure that the generated test set 
can explore new fault-inducing data flows (i.e., 
$N_{i}(I+E(\theta))$ which has a large difference with 
$N_{i}(I))$).
\fuzzer then feeds the generated 
$I+E(\theta')$ to \calculator 
for computing $N_{i}(I+E(\theta'))-N_{i}(I))$, and thus \xxx 
enters another iteration of the testing.

\section{\xxx's runtime}\label{sec:algo}

\subsection{The challenge in designing \xxx's workflow}\label{sec:algo:challenge}

One major novelty of 
\xxx is a new workflow to compute 
test coverage metric, by computing $\hat{ND_{i}}$ for 
all of a DLS's neurons 
(\S\ref{sec:back:testing} and \S\ref{sec:overview}). 
However, realising such a workflow 
is challenging, because  
$\hat{ND_{i}}$ for all of a DLS's neurons may cause \xxx 
unscalable to a DLS's size. 
\begin{table}
    \tiny
    \centering
    \begin{tabular}{|c|c|c|c|}
    \hline
    Dataset \textbf{(I)}      &  Description                                                                & DNN \textbf{(M)}     & \begin{tabular}[c]{@{}l@{}}Number of \\ neurons \end{tabular} \\ \hline
    \multirow{3}{*}{MNIST}    & \multirow{3}{*}{\begin{tabular}[c]{@{}l@{}}Hand-written digit \\ classification\end{tabular}} & LeNet-1~\cite{lecun1998gradient,deng2012mnist}                              & 7206               \\
                              &                                                    & LeNet-4~\cite{lecun1998gradient,deng2012mnist}                               & 69362                  \\
                              &                                                    & LeNet-5~\cite{lecun1998gradient,deng2012mnist}                               & 107786                   \\ \hline
    Contagio       & \begin{tabular}[c]{@{}l@{}}Malware classification \\ in PDF Files\end{tabular}                & $<$200,200$>$~\cite{pei2017deepxplore}                                           & 35410             \\ \hline
    Drebin                    & \begin{tabular}[c]{@{}l@{}}Malware classification \\ in Android apps\end{tabular}             & $<$200,10$>$~\cite{pei2017deepxplore,grosse2016adversarial}     & 15230             \\ \hline
    \multirow{2}{*}{ImageNet} & \multirow{2}{*}{\begin{tabular}[c]{@{}l@{}}General Image \\ classification\end{tabular}}      & VGG-19~\cite{simonyan2014very}                                               & 14888             \\
                              &                                                    & ResNet-50~\cite{he2016deep}                                                  & 16168             \\ \hline
    Udacity                   & \begin{tabular}[c]{@{}l@{}}Road condition \\ classification\end{tabular}                      & DAVE-2~\cite{zhou2016dave}                                                   & 1560              \\ \hline
    \multirow{2}{*}{Cifar10}  & \multirow{2}{*}{\begin{tabular}[c]{@{}l@{}}General Image \\ classification\end{tabular}}      & ResNet56~\cite{he2016deep}                                                   & 532490            \\
                              &                                                   & DenseNet121~\cite{huang2017densely}                                          & 563210            \\ \hline
    \end{tabular}
    \caption{Datasets and DNNs for evaluating \xxx, which covers the complete set of 
    datasets evaluated by baselines.}
    \label{tab:dataset}
\end{table}

Computing 
$\hat{ND_{i}}$ for just one neuron of a DLS is 
already computationally 
expensive, 
because it requires \xxx's \coverage 
to run MCMC on  
thousands of sensitivity samples 
(\S\ref{sec:overview}). 
Hence, for DLSs of large 
sizes (e.g., ResNet56 which consists of more than five 
hundred thousand neurons, see Table~\ref{tab:dataset}), the strawman approach of brutal force computing 
would require more than hours to conduct 
the testing, inefficient 
compared to the existing techniques~\cite{pei2017deepxplore, 
ma2018deepGauge, 
gerasimou2020importance, 
kim2019guiding}, which only takes several 
minutes to complete testing. 
On the other hand, 
computing $\hat{ND_{i}}$ for only a 
subset of a DLS's neurons may cause \xxx 
to miss substantial faults. 
Because $\hat{ND_{i}}$ of different neurons 
have different convergence rate ($\hat{ND_{i}}$'s convergence rate of a DLS 
neuron is essential for \xxx to compute 
test coverage, see \S\ref{sec:overview}), 
computing $\hat{ND_{i}}$ for a subset of a DLS's neurons may cause 
\xxx to compute test coverage incorrectly (e.g., stop testing 
too soon or too late). 

To tackle this challenge, we present \sampler, 
a sampling technique used by \xxx's \coverage 
to compute $\hat{ND_{i}}$ for a subset of a 
DLS's neurons, in order to precisely 
approximate the $\hat{ND_{i}}$'s convergence rate of 
all the DLS's neurons. Our observation behind \sampler 
is that $\hat{ND_{i}}$'s convergence rate 
is inversely 
proportional to $\hat{ND_{i}}$'s variance (the larger the variance, 
the slower the convergence 
rate, as pointed out by a classic statistic theory called 
Chebyshev's inequality~\cite{marshall1960multivariate}). 
Hence, for a DLS of any size, \xxx samples a constant number 
(by default one thousand, see \S\ref{sec:algo:algo}) of neurons 
for computing $\hat{ND_{i}}$ based on the variance of 
$N_{i}(I+E(\theta))-N_{i}(I))$ values of each neuron. 

\begin{algorithm}[tb]\label{algo}
    \small
    \DontPrintSemicolon
    \SetNoFillComment
    \SetKwInput{Vars}{Variables}
    \caption{\xxx's entire workflow}\label{algo}
    \KwIn{
     $\{I\}$: raw inputs,
     $M$: DNN to be tested, 
     $n$: the number of neurons of $M$,
     $E(\theta)$: User-specified perturbation,
     $t$: threshold value for MCMC to determine convergence,
     $k$: sample size for \sampler,
     $c$: desired sensitivity coverage 


    }
    \SetKwProg{Fn}{Function}{:}{}
    \SetKwFunction{frr}{rr}
    \SetKwFunction{fsampler}{sampler}

    \Fn{\xxx($\{I\}$,$E(\theta)$,$M$,$n$)}{
        Initialise List $sensitivitySample$\label{algo:phase1:start};
        Initialise List $Faults$ \;
        \While{True}{
            \tcc{\calculator (Phase 1) begins}
            \For { each $I \in \{I\}$ } {
                $\{{N_{i}(I)}\}_{i=1}^{n} \leftarrow M(\{I\})$'s activtion values \; \label{algo:compav:start}
                $\{{N_{i}(I+E(\theta))}\}_{i=1}^{n} \leftarrow M(I+E(\theta))$'s activtion values \;\label{algo:compav:end}
                Initialise Array $\{Diff_{i}\}_{i=1}^{n}$\;
                \For { each $j \in $ [1,...,n] } {\label{algo:diffneu:start}
                    $ Diff_{j} = |N_{j}(I+E(\theta)) - N_{j}(I)|$\;
                }   
                $sensitivitySample$.insert($Diff$)\; \label{algo:diffneu:end}
                \If{$M(I) \neq M(I+E(\theta))$} {
                    $Faults$.insert($I$,$E(\theta)$)\;  
                }
            }\label{algo:phase1:end}
            \tcc{\coverage (Phase 2) begins}
            $\{Diff_{i}\}_{i=1}^{k} \leftarrow$ sensitivitySampler($sensitivitySample$,$k$)\;\label{algo:phase2:start}
            $\{\hat{ND_{i}}\}_{i=1}^{k}$ $\leftarrow$ MCMC($\{Diff_{i}\}_{i=1}^{k}$)\;\label{algo:phase2:mcmc}
            sensitivityCoverage $\leftarrow$ proportion of $\{\hat{ND_{i}}\}$ that has MCSE $== 0.0$\;\label{algo:phase2:end}
            \tcc{\fuzzer (Phase 3) begins}
            \If{sensitivityCoverage $<  c$} {
                $UncovNeuron \leftarrow$ unconverged neuron \;
                obj $=$ maximizer($UncovNeuron$) \;\label{algo:phase3:start}
                $I+E(\theta') \leftarrow$ gradientAscent(obj,$I+E(\theta)$) \;\label{algo:phase3:end}

            }
            \Else{
                break \;\label{algo:break}
            }

        }
        \KwRet 
    }

    \Fn{sensitivitySampler($sensitivitySample$,$k$)}{\label{algo:sampler:start}
        Initialise Array $\{variance_{i}\}_{i=1}^{n}$\;
        \For { each $j \in$ [1,...,n]) } {
            $Variance_{j}$ = compute\_variance($sensitivitySample[j]$)\;
        }
        $D_{sort}$ = sort $\{Variance_{i}\}_{i=1}^{n}$ based on variance\;
        \For { each $j \in$ [1,...,k]) } {
            {\color{\curr}$selected_{j} = variance_{j*\frac{|D_i|}{k}}$\;}
        }
        \KwRet $\{selected_{i}\}_{i=1}^{k}$\;\label{algo:sampler:end}
    }
\end{algorithm}

\subsection{\xxx's algorithm}\label{sec:algo:algo}

Algorithm~\ref{algo} shows the 
algorithm of \xxx, including \xxx's three 
components (\S\ref{sec:overview}): 
\calculator (Phase 1), \coverage (Phase 2) and 
\fuzzer (Phase 3), as well 
as \sampler (\S\ref{sec:algo:challenge}). 
When $M$, $\{I\}$, and $E(\theta)$ are fed into \xxx, 
\xxx starts testing process. Specifically, 
for $I$ in $\{I\}$, 
\xxx's \calculator 
(line~\ref{algo:phase1:start}-\ref{algo:phase1:end}) 
first computes $N_{i}(I)$ and $N_{i}(I+E(\theta))$ 
(line~\ref{algo:compav:start}-\ref{algo:compav:end}), 
which are defined in \S\ref{sec:back:prob}. 
\calculator then computes the 
difference (denoted as $D_i$) between  
$N_{i}(I)$ and $N_{i}(I+E(\theta))$ (line~\ref{algo:diffneu:start}-
\ref{algo:diffneu:end}). $D_i$ are  
\textit{sensitivity samples} (\S\ref{sec:overview}). Meanwhile, 
\calculator reports faults to the developer for any 
$M(I)$ and $M(I+E(\theta))$ with different DLS's outputs. 

Then, \xxx's \coverage 
infers $\hat{ND_{i}}$ of the DLS's neurons based on $D_i$ 
(line~\ref{algo:phase1:start}-~\ref{algo:phase2:end}). Specifically, \sampler computes the variance of each item in $D_i$, and sort these items according to their variance. 
Then \sampler selects $k$ (by default one thousand) items from all the items in an evenly-spaced manner (i.e., select the item for every $\frac{|D_i|}{k}$) items). The default value 
of $k$ (i.e., one thousand) is selected 
based on Mann-Witney statistic~\cite{birnbaum2020use}, 
which points out that one thousand samples from a 
sorted list is sufficient to accurately 
approximate all the remaining values  
in the sorted list. 


{\color{\curr}
After \sampler computes $\hat{ND_{i}}$ 
for each sampled neurons (line~\ref{algo:phase2:mcmc}), 
\sampler computes \textit{sensitivity convergence} (the test 
coverage of \xxx) as the number of 
converged $\hat{ND_{i}}$ over $k$ 
(line~\ref{algo:phase2:end}). Specifically, the convergence 
criterion is whether Monte Carlo Standard Error (MCSE) equals 
0.0, a standard criterion which indicates the probability of convergence  
identified by 
MCMC~\cite{raychaudhuri2008introduction,
geyer1992practical}. 
By having MCSE equal 0.0, 
each batch of samples drawn by MCMC have almost 
identical probability distributions. We also 
evaluated the choice of MCSE value and our evaluation 
result (Figure~\ref{fig:sensitivity}b) 
shows that MCSE equal 0.0 allowed \sampler 
to precisely infer $\hat{ND_{i}}$'s convergence 
rate of all neurons. }
After \sampler computed \textit{sensitivity convergence}, 
\xxx's \fuzzer (line~\ref{algo:phase3:start}-
line~\ref{algo:phase3:end}) performs fuzzing in order to explore 
the unconverged $\hat{ND_{i}}$ (i.e., coverage-guided fuzzing). 

\textbf{Analysis of  
\xxx's fault detection capability}\label{sec:analysis}.
\xxx has strong fault 
detection capability because
\xxx theoretically can explore all fault-inducing 
data flows at high probability (95\%). We 
first explain how we derive 
this theoretical property of \xxx. As discussed 
in \S\ref{sec:back:mcmc}, we  
summarize two existing 
theories~\cite{NEURIPS2018_69386f6b,
long2018pde,robert1998discretization,
zheng2016improving} and conclude that 
when random noise 
is present on a DLS's input, the variation of 
a DLS's activation values (i.e., $N_{i}(I+E) - N_{i}(I)$ 
defined in \S\ref{sec:back:testing}) often follows a normal 
distribution (i.e., $ND_{i}$ 
defined in \S\ref{sec:back:testing}). Hence, 
\xxx examines whether the generated test set explores 
a full coverage of the fault-inducing data flows 
(i.e., $N_{i}(I+E)$ which has large difference with 
$N_{i}(I)$) by examining 
whether a normal distribution can 
be inferred from the $N_{i}(I+E) - N_{i}(I)$ values 
associated to the test set (i.e., whether the test set 
explores a full coverage of 
$N_{i}(I+E) - N_{i}(I)$ values). 

Note that even when distribution 
inferred from the $N_{i}(I+E) - N_{i}(I)$ values 
(i.e., $\hat{ND_{i}}(I+E)$) converges to 
a normal distribution, \xxx is not 100\% guaranteed 
to identify all fault-inducing data flows. 
It is because the $N_{i}(I+E) - N_{i}(I)$ 
values can coincidentally form a normal distribution 
different from $ND_{i}$. 
Hence, \xxx probabilistically (rather than 
deterministically 100\%) identifies a full coverage 
of fault-inducing data flows, 
and the corresponding probability 
depends on the confidence interval (by default 
95\%) adopted by \xxx when 
inferring $ND_{i}$ (\S\ref{sec:overview}).

After we justify the reason why 
\xxx theoretically can explore all fault-inducing 
data flows at high probability (95\%), we show that 
this theoretical property of \xxx allows \xxx to 
have strong fault detection capability. As discussed 
in \S\ref{sec:algo:algo},
for any DLS having more number 
of faults (i.e., has lower accuracy), the 
DLS's $\hat{ND_{i}}$ often has greater variance. This implies that 
these DLSs require
\xxx to generate more $I+E(\theta)$  
to identify the DLS's $\hat{ND_{i}}$ 
(any statistical distribution with greater variance 
requires more samples to identify the distribution, 
according to 
Chebyshev's inequality~\cite{marshall1960multivariate}). Since with  
more $I+E(\theta)$, \xxx can detect 
more faults (the number of $I+E(\theta)$ generated by \xxx is 
the upper limit of the number of faults can be detected by \xxx). 
Hence, \xxx detects more faults from a DLS which has lower 
accuracy (i.e., \xxx has strong fault detection capability).

\section{Implementation}

We implemented \xxx using 
PyTorch 1.8.1~\cite{paszke2019pytorch}. 
\xxx's implementation consists 
of around 9,791 lines of Python code. 
We adopted \textsc{pymc3}~\cite{Salvatier2016}, 
a popular Python package which realises MCMC, to 
perform MCMC in \xxx. Specifically, \xxx's 
\calculator (\S\ref{sec:overview}) calls 
\textsc{pymc3.Data} to load the 
samples of sensitivity, and constructs 
the prior distributions by calling 
\textsc{pymc3.Normal}. Then, 
\xxx's \coverage calls 
\textsc{pymc3.sample} to perform MCMC. 
\coverage examines the convergence rate (i.e., MCSE, 
\S\ref{sec:algo}) by \textsc{arviz.summary}~\cite{arviz_2019}, 
a popular Python package for analysis of Bayesian models. 
Finally, \xxx's \fuzzer is realised with textsc{Autograd}~\cite{maclaurin2016modeling} library, 
which efficiently 
performs gradient ascent on $\hat{ND_{i}}$(\S\ref{sec:algo}). 

\begin{table*}[!th]

    \begin{centering}
    
    \begin{tabular}{|l|lllll|lllll|}
    \hline 
    \multirow{3}{*}{{\small{}Dataset(DNN)}} & \multicolumn{5}{c|}{{\small{}CW}} & \multicolumn{5}{c|}{{\small{}FGSM}}\tabularnewline
        & \multirow{2}{*}{{\small{}{\small{}\xxx{}}}} & {\small{}Deep-} & {\small{}Deep-} & {\small{}Deep-} & {\small{}Surprise} & \multirow{2}{*}{{\small{}{\small{}\xxx{}}}} & {\small{}Deep-} & {\small{}Deep-} & {\small{}Deep-} & {\small{}Surprise}\tabularnewline
        &  & {\small{}Xplore} & {\small{}Gauge} & {\small{}Import.} & {\small{}Adequacy} &  & {\small{}Xplore} & {\small{}Gauge} & {\small{}Import.} & {\small{}Adequacy}\tabularnewline
    \hline 
    {\small{}MNIST (LeNet-1)} & \textcolor{teal}{0.79} & \textcolor{brown}{-0.58} & 0.25 & 0.08 & \textcolor{brown}{-0.21} & \textcolor{teal}{0.77} & 0.21 & 0.35 & \textcolor{brown}{-0.35} & 0.33\tabularnewline
    {\small{}MNIST (LeNet-4)} & \textcolor{teal}{0.72} & \textcolor{brown}{-0.88} & 0.01 & \textcolor{brown}{-0.62} & \textcolor{brown}{-0.5} & \textcolor{teal}{0.74} & 0.26 & 0.34 & 0.24 & \textcolor{brown}{-0.18}\tabularnewline
    {\small{}MNIST (LeNet-5)} & \textcolor{teal}{0.71} & 0.35 & \textcolor{brown}{-0.41} & \textcolor{brown}{-0.17} & \textcolor{teal}{0.74} & \textcolor{teal}{0.83} & \textcolor{brown}{-0.02} & \textcolor{brown}{-0.02} & 0.18 & \textcolor{brown}{-0.05}\tabularnewline
    {\small{}Contagio ($<$200,200$>$)} & \textcolor{teal}{0.71} & 0.33 & \textcolor{brown}{-0.07} & 0.25 & \textcolor{brown}{-0.87} & \textcolor{teal}{0.77} & 0.71 & \textcolor{brown}{-0.04} & 0.1 & 0.38\tabularnewline
    {\small{}Drebin ($<$200, 10$>$)} & \textcolor{teal}{0.70} & 0.11 & 0.18 & \textcolor{teal}{0.78} & \textcolor{brown}{-0.37} & \textcolor{teal}{0.81} & \textcolor{brown}{-0.06} & 0.08 & \textcolor{brown}{-0.19} & 0.2\tabularnewline
    {\small{}ImageNet (VGG-19)} & \textcolor{teal}{0.79} & \textcolor{brown}{-0.77} & \textcolor{brown}{-0.07} & 0.12 & \textcolor{brown}{-0.29} & \textcolor{teal}{0.86} & 0.3 & \textcolor{brown}{-0.62} & \textcolor{brown}{-0.01} & \textcolor{teal}{0.74}\tabularnewline
    {\small{}ImageNet (ResNet-50)} & \textcolor{teal}{0.71} & 0.4 & 0.33 & \textcolor{brown}{-0.06} & \textcolor{brown}{-0.09} & \textcolor{teal}{0.74} & 0.3 & 0.1 & \textcolor{brown}{-0.05} & 0.1\tabularnewline
    {\small{}Udacity (DAVE-2)} & \textcolor{teal}{0.72} & \textcolor{brown}{-0.16} & 0.31 & \textcolor{teal}{0.76} & 0.21 & \textcolor{teal}{0.84} & \textcolor{teal}{0.77} & 0.4 & 0 & \textcolor{brown}{-0.08}\tabularnewline
    {\small{}Cifar10 (ResNet56)} & \textcolor{teal}{0.86} & 0.35 & \textcolor{brown}{-0.36} & 0.4 & 0.21 & \textcolor{teal}{0.85} & 0.37 & 0.01 & \textcolor{brown}{-0.63} & 0.08\tabularnewline
    {\small{}Cifar10 (DenseNet121)} & \textcolor{teal}{0.88} & 0.25 & \textcolor{brown}{-0.7} & 0.05 & \textcolor{brown}{-0.01} & \textcolor{teal}{0.78} & \textcolor{brown}{-0.51} & \textcolor{teal}{0.72} & 0.33 & 0.39\tabularnewline
    \hline 
    \multirow{3}{*}{{\small{}Dataset(DNN)}} & \multicolumn{5}{c|}{{\small{}PGD}} & \multicolumn{5}{c|}{{\small{}GAUSSIAN}}\tabularnewline
        & \multirow{2}{*}{{\small{}{\small{}\xxx{}}}} & {\small{}Deep-} & {\small{}Deep-} & {\small{}Deep-} & {\small{}Surprise} & \multirow{2}{*}{{\small{}{\small{}\xxx{}}}} & {\small{}Deep-} & {\small{}Deep-} & {\small{}Deep-} & {\small{}Surprise}\tabularnewline
        &  & {\small{}Xplore} & {\small{}Gauge} & {\small{}Import.} & {\small{}Adequacy} &  & {\small{}Xplore} & {\small{}Gauge} & {\small{}Import.} & {\small{}Adequacy}\tabularnewline
    \hline 
    {\small{}MNIST (LeNet-1)} & \textcolor{teal}{0.72} & \textcolor{brown}{-0.89} & 0.01 & 0.39 & 0.23 & \textcolor{teal}{0.7} & 0.2 & \textcolor{brown}{-0.21} & \textcolor{brown}{-0.06} & \textcolor{teal}{0.71}\tabularnewline
    {\small{}MNIST (LeNet-4)} & \textcolor{teal}{0.77} & \textcolor{brown}{-0.03} & \textcolor{brown}{-0.54} & 0.37 & 0.19 & \textcolor{teal}{0.72} & 0.18 & \textcolor{brown}{-0.07} & \textcolor{brown}{-0.6} & \textcolor{brown}{-0.68}\tabularnewline
    {\small{}MNIST (LeNet-5)} & \textcolor{teal}{0.71} & 0.01 & \textcolor{brown}{-0.04} & \textcolor{brown}{-0.23} & \textcolor{teal}{0.76} & \textcolor{teal}{0.89} & \textcolor{brown}{-0.17} & \textcolor{brown}{-0.09} & \textcolor{teal}{0.75} & 0\tabularnewline
    {\small{}Contagio ($<$200,200$>$)} & \textcolor{teal}{0.78} & \textcolor{brown}{-0.13} & 0.3 & 0.23 & \textcolor{brown}{-0.09} & \textcolor{teal}{0.77} & \textcolor{brown}{-0.63} & 0.18 & 0.37 & 0.28\tabularnewline
    {\small{}Drebin ($<$200, 10$>$)} & \textcolor{teal}{0.71} & 0.4 & \textcolor{brown}{-0.02} & \textcolor{brown}{-0.1} & \textcolor{brown}{-0.57} & \textcolor{teal}{0.82} & 0.26 & \textcolor{brown}{-0.32} & 0.36 & 0.11\tabularnewline
    {\small{}ImageNet (VGG-19)} & \textcolor{teal}{0.75} & 0.09 & 0.1 & \textcolor{brown}{-0.13} & \textcolor{brown}{-0.19} & \textcolor{teal}{0.84} & \textcolor{teal}{0.79} & 0.28 & 0.1 & \textcolor{brown}{-0.03}\tabularnewline
    {\small{}ImageNet (ResNet-50)} & \textcolor{teal}{0.78} & \textcolor{brown}{-0.09} & \textcolor{brown}{-0.56} & 0.02 & \textcolor{brown}{-0.05} & \textcolor{teal}{0.86} & \textcolor{brown}{-0.16} & \textcolor{brown}{-0.17} & \textcolor{brown}{-0.09} & \textcolor{brown}{-0.17}\tabularnewline
    {\small{}Udacity (DAVE-2)} & \textcolor{teal}{0.71} & 0.31 & 0.07 & \textcolor{brown}{-0.71} & \textcolor{brown}{-0.74} & \textcolor{teal}{0.8} & 0.4 & 0.34 & \textcolor{brown}{-0.54} & 0.08\tabularnewline
    {\small{}Cifar10 (ResNet56)} & \textcolor{teal}{0.8} & \textcolor{brown}{-0.1} & 0.09 & \textcolor{brown}{-0.08} & 0.36 & \textcolor{teal}{0.95} & \textcolor{brown}{-0.57} & 0.11 & 0.06 & 0.35\tabularnewline
    {\small{}Cifar10 (DenseNet121)} & \textcolor{teal}{0.86} & \textcolor{brown}{-0.04} & 0.25 & \textcolor{brown}{-0.56} & \textcolor{brown}{-0.73} & \textcolor{teal}{0.91} & 0.22 & \textcolor{brown}{-0.14} & 0.31 & 0.4\tabularnewline
    \hline 
    \end{tabular}
    \par\end{centering} 
    
    \caption{Correlation between the number 
    of faults identified by DLS testing and 
    the error rate of a DLS. 
    Correlation larger than 0.7 (colored in green) 
    is considered strong~\cite{harel2020neuron}.
    Correlation less than 0.0 is colored in red.}
    \label{tab:corr}
\end{table*}
\section{Evaluation}\label{sec:eval}
\begin{table}
    \center
    \begin{tabular}{|l|l|l|l|l|l|}
    \hline
    Variables (Perturbations)        & \multicolumn{5}{c|}{Magnitude Value}                            \\ \hline
    $c$ Confidence (CW~\cite{carlini2017towards}) & 10   & 20   & 30                        & 40   & 50   \\ \hline
    $\epsilon$ (FGSM~\cite{goodfellow2014explaining})    & 0.1  & 0.2  & 0.3                       & 0.4  & 0.5  \\ \hline
    $\epsilon$ (PGD~\cite{madry2017towards})     & 0.1  & 0.2  & 0.3                       & 0.4  & 0.5  \\ \hline
    $\sigma$ (Guassian noise)  & 0.01 & 0.02 & \multicolumn{1}{c|}{0.03} & 0.04 & 0.05 \\ \hline
    \end{tabular}
    \caption{Perturbations adopted in our evaluation.}
    \label{tab:var}
\end{table}

\subsection{Experiment Setup}\label{sec:setup}

Our evaluation was done on a
a computer equipped with twenty CPU cores and four
Nvidia RTX2080TI graphic cards. 
Inspired by~\cite{HarelCanada2020IsNC} that analyze the correlation of increases in neuron coverage and ASR, we compared \xxx with
the state-of-the-art DLS testing techniques: 
DeepeXplore\cite{pei2017deepxplore}, 
DeepGauge\cite{ma2018deepGauge}, DeepImportance\cite{gerasimou2020importance}, and Surprise
Adequacy\cite{kim2019guiding}. During the testing process, we will mutate test cases with perturbations in~\ref{tab:var} and require the NC~(DeepXplore), KMNC~(DeepGauge), IDC~(DeepImportance), and LSA~(Surprise Adequacy) to select test cases to increases the coverage rate. For NC, we set the threshold as 0.5, For KMNC, we follow the configuration~\cite{ma2018deepGauge} and set k as 1,000. For LSA, we set the upper bound as 2,000. For IDC, we set the $m$ as 12.

Table~\ref{tab:dataset} shows our 
evaluated datasets and models, which 
covers a complete set of six datasets 
evaluated by four baselines. 
To be fair and comprehensive, our evaluated models covered \textbf{all} models evaluated by four notable DL testing works~\cite{pei2017deepxplore,
gerasimou2020importance,
kim2019guiding,ma2018deepGauge} that have been deployed in Mindspore security framework.
The models we evaluated include VGG~\cite{simonyan2014very}, ResNet~\cite{he2016deep}, DenseNet~\cite{huang2017densely}, etc. We believe the architecture of these models covers all basic building blocks (such as convolution layers and residual connections) of most modern real-world deployed AI applications. 
These DNN models all achieve high precision. Hence, finding faults (i.e., 
a DNN's incorrect outputs) 
from these well-studied and well-tested DNN will be valuable.

We choose the correlation metric~\cite{harel2020neuron,li2019structural} as the main evaluation metric to compare the \fdc between \xxx and the baselines. The value of the correlation metric is between -1.0 and 1.0 and the latest study~\cite{harel2020neuron} points out that a high correlation (e.g., 0.7) implies strong \fdc. We also evaluated the increased accuracy of DLS after retraining the DLS with the detected faults, to evaluate the quality of faults detected by a DLS testing technique. Finally, we evaluated the test time cost for \xxx and the baselines to achieve 100\% test coverage\footnote{In this paper, we follow the Mindspore fuzzing testing setup for our experiments, where the testing will be stopped when the coverage reaches 100\%.}.

To study \xxx and the baselines' \fdc, 
we applied four well-studied
perturbations 
(CW~\cite{carlini2017towards}, 
FGSM~\cite{goodfellow2014explaining}, 
PGD~\cite{madry2017towards}, and 
Gaussian noise) to each 
dataset of the 
evaluated DNN models in Table~\ref{tab:dataset}. 
These perturbations are 
evaluated by \xxx's 
baselines~\cite{ma2018deepGauge,gerasimou2020importance} 
because these perturbations are common 
in real world DLS applications. Hence, faults detected 
from a DLS under these perturbations 
greatly promote the reliability of a DLS. 

The values of the magnitude of these 
perturbations in our evaluation are listed 
in Table~\ref{tab:var}. We follow 
\xxx's baselines to set these values~\cite{pei2017deepxplore,
ma2018deepGauge,
gerasimou2020importance,
kim2019guiding}. The evaluation questions are as follows:

\begin{tightenum}

    \item[\S\ref{sec:eval:fdc}:] How is \xxx's \fdc compared to the baselines?
    
    \item[\S\ref{sec:eval:acc}:] How is the improvement on a DLS's accuracy attained by \xxx?
    
    \item[\S\ref{sec:eval:perf}:] How is \xxx's efficiency compared to baselines?

    \item[\S\ref{sec:eval:sens}:] What are the factors that affect \xxx's \fdc?
    
    
    
    \item[\S\ref{sec:eval:limit}:] What are \xxx's limitations? 
    
\end{tightenum}



\subsection{\xxx's Fault Detection Capability Result}\label{sec:eval:fdc}

We first investigate the \fdc of \xxx and baselines. 
{\color{\note}
Specifically, we perturbed raw inputs (e.g., road condition images) with various noise intensities as shown in Table~\ref{tab:var}. Then, for each set of inputs perturbed by the same noise intensity (e.g., PGD with epsilon as 0.1), we fed these perturbed inputs into a DLS and computed the DLS’s error rate (i.e., the proportion of the perturbed inputs being wrongly classified by the DLS) and the number of faults detected by \xxx until \xxx’s coverage metric reached 100\%. Finally, the \fdc is computed as the correlation between all pairs of a DLS’s error rate and the number of faults detected (e.g., the pairs of DLS’s error rate and the number of faults detected corresponding to PGD’s epsilon as 0.1,0.2,...0.5 as shown in Table~\ref{tab:var}). Then, the inference accuracy of these 
DNNs and the number of faults detected by \xxx or the 
baselines are used to compute the \fdc of \xxx and 
the baselines.
}

Table~\ref{tab:corr} shows the \fdc of \xxx{} and the baselines on all datasets. 
\xxx's correlation was larger than 
0.7 for all the evaluated 
datasets, while baselines rarely achieved 0.7 correlation or even 
had negative correlation (i.e., detected fewer faults for DNN 
which has lower accuracy). It is because the
baselines tested a DLS deterministically, 
while \xxx tested a DLS 
probabilistically (\S\ref{sec:back:prob}). We further inspect why \xxx achieved strong 
\fdc in Figure~\ref{fig:mcmcmean}b. 
As discussed in \S\ref{sec:algo:algo}, the condition for 
\xxx to achieve strong \fdc is that  
\xxx precisely computes $\hat{ND_{i}}$, which guides \xxx 
to attain full fault coverage. Hence, we inspect 
whether \xxx precisely computes $\hat{ND_{i}}$, 
by inspecting the consistency of $\hat{ND_{i}}$'s means computed by 
\xxx: \xxx is supposed to compute 
the same $\hat{ND_{i}}$'s mean for the same DLS on different 
sets of test inputs because theoretically, 
there is only one $\hat{ND_{i}}$ for each 
DLS's neuron for the same set of test inputs 
(\S\ref{sec:back:prob}).  

Figure~\ref{fig:mcmcmean}b shows that \xxx 
computed $\hat{ND_{i}}$'s mean with less than 10\% deviation, so 
according to conventional standards 
in statistics~\cite{harding2014standard}, 
\xxx successfully computed $\hat{ND_{i}}$. 
\xxx can do this because theoretically, 
$\hat{ND_{i}}$ could be 
identified by Bayesian analysis, and 
\xxx's MCMC component could 
identify it. Figure~\ref{fig:efficiency}b 
also confirms our observation that greater 
sensitivity ($N_{i}(I+E(\theta)) - N_{i}(I)$) 
values resulted in higher number of faults.

\subsection{Increase in accuracy obtained by \xxx}\label{sec:eval:acc}

\begin{figure}
    \begin{minipage}{0.4\textwidth}
        \centering
        \includegraphics[width=1.0\textwidth]{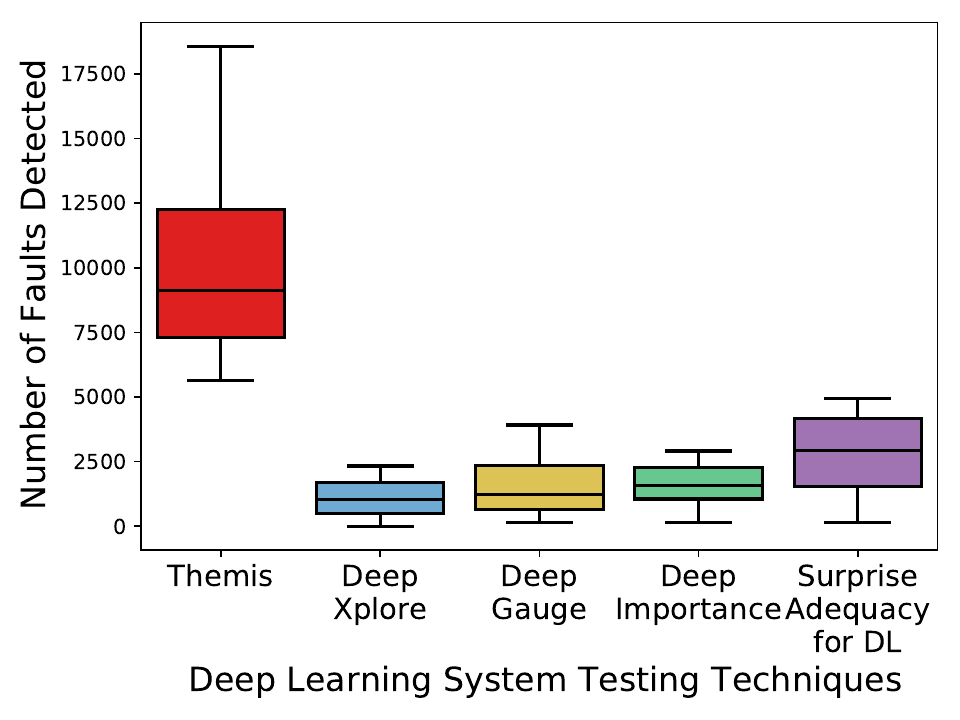}
    \end{minipage}
    \begin{minipage}{0.4\textwidth}
        \centering
        \includegraphics[width=1.0\textwidth]{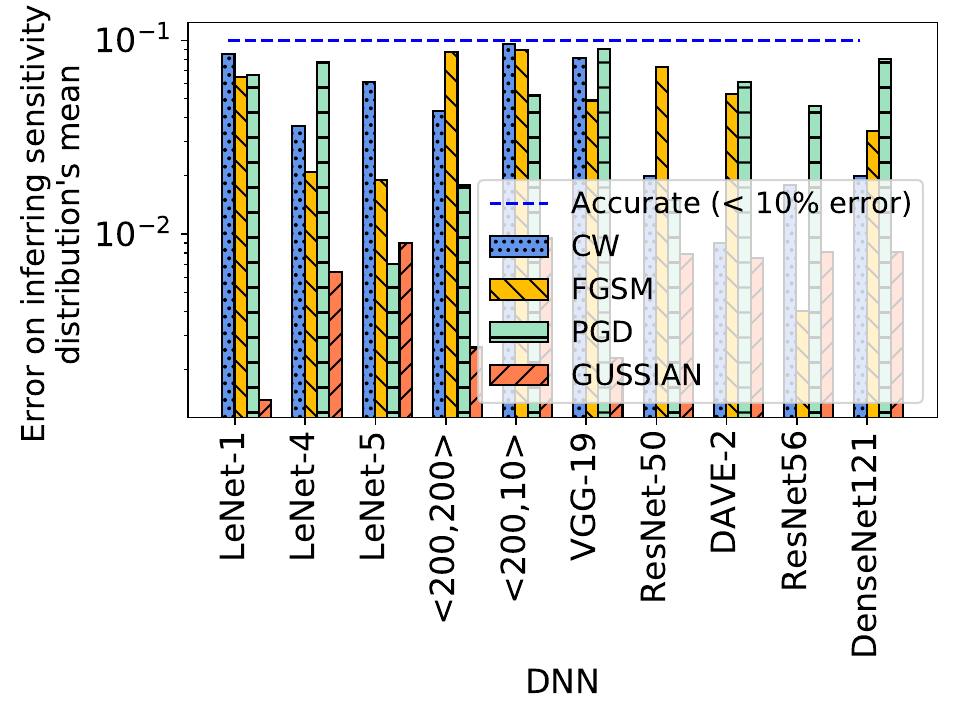}
    \end{minipage}
    \caption{Above(a): Number of faults detected by \xxx and the baselines when their test coverage metric reaches 100\%.
    Bottom(b) The mean of $\hat{ND_{i}}$ inferred by \xxx for each evaluated DNN. \xxx is considered 
    accurate in inferring the mean if the variation of the mean is 
    below 10\%~\cite{harding2014standard}
    .}
    \label{fig:mcmcmean}
\end{figure}

One main purpose of DLS testing is to retrain a DLS 
with detected faults to increase the DLS's 
accuracy. Figure~\ref{fig:intro} shows  
the increase in DLS's accuracy 
brought by \xxx and the baselines. 
Overall, \xxx increased the DLS's accuracy 
higher than baselines for all datasets. 
The main reason is 
that \xxx identified more faults than the baselines 
(Figure~\ref{fig:mcmcmean}a). {\color{\note}Specifically, in our evaluation, we let both \xxx and the baselines keep generating inputs until their coverage metric reached 100\%. By doing so, \xxx on average generated 21399 test inputs, DeepXplore~\cite{pei2017deepxplore} generated 3774 test inputs, DeepGuage~\cite{ma2018deepGauge} generated 18501 test inputs, DeepImportance~\cite{gerasimou2020importance} generated 20052 test inputs, and Surprise Adequacy~\cite{kim2019guiding} generated 26757 test inputs. Overall, the test inputs generated by \xxx consisted of much more faults than the baselines.}






We found that \xxx increased the accuracy of Cifar10 (densenet121) under PGD perturbation the most (i.e., \vggacc). It is because densenet121 is vulnerable to PGD perturbation~\cite{rahnama2020robust}, so it suffered from the greatest accuracy loss under this perturbation. Since \xxx detected more faults from DLS which had lower accuracy (Table~\ref{tab:corr}), \xxx detected more faults for this DNN than other DNNs, which allowed \xxx to increase the DNN's accuracy more than \xxx did on the other DNNs. This also implies \xxx is valuable to real-world DLSs, which are often trained with limited datasets and 
have moderate accuracy (especially on safety-critical tasks, see \S\ref{sec:intro}). 

\begin{table*}[]

    \noindent \begin{centering}
    \begin{tabular}{|l|llll|llll|}
    \hline 
    \multirow{3}{*}{{\footnotesize{}Dataset(DNN)}} & \multicolumn{4}{c|}{{\footnotesize{}with \sampler}} & \multicolumn{4}{c|}{{\footnotesize{}without \sampler}}\tabularnewline
    \cline{2-9} \cline{3-9} \cline{4-9} \cline{5-9} \cline{6-9} \cline{7-9} \cline{8-9} \cline{9-9} 
     & {\footnotesize{}Total} & {\footnotesize{}Sensitivity} & {\footnotesize{}Sensitivity} & {\footnotesize{}{\footnotesize{}Sensitivity} Fuzzer} & {\footnotesize{}Total} & {\footnotesize{}Sensitivity} & {\footnotesize{}Calculator} & {\footnotesize{}Sensitivity Fuzzer}\tabularnewline
     & {\footnotesize{}Test Time} & {\footnotesize{}Calculator} & {\footnotesize{}Coverage} & {\footnotesize{}(iterations)} & {\footnotesize{}Test Time} & {\footnotesize{}Calculator} & {\footnotesize{}Estimator} & {\footnotesize{}(iterations)}\tabularnewline
    \hline 
    {\footnotesize{}MNIST (LeNet-1)} & \textcolor{teal}{160} & 27 & 24 & 109 (5) & 11171 & 24 & 2006 & 9141 (5)\tabularnewline
    {\footnotesize{}MNIST (LeNet-4)} & \textcolor{teal}{143} & 22 & 11 & 110 (10) & 2434 & 23 & 217 & 2194 (11)\tabularnewline
    {\footnotesize{}MNIST (LeNet-5)} & \textcolor{teal}{208} & 59 & 34 & 115 (4) & 1704 & 56 & 379 & 1269 (4)\tabularnewline
    {\footnotesize{}Contagio ($<$200,200$>$)} & \textcolor{teal}{201} & 49 & 9 & 143 (15) & 14320 & 50 & 856 & 13414 (16)\tabularnewline
    {\footnotesize{}Drebin ($<$200,10$>$)} & \textcolor{teal}{211} & 27 & 17 & 167 (10) & 12690 & 27 & 1139 & 11524 (12)\tabularnewline
    {\footnotesize{}ImageNet (VGG-19)} & \textcolor{teal}{202} & 49 & 23 & 130 (6) & 10273 & 50 & 1533 & 8690 (6)\tabularnewline
    {\footnotesize{}ImageNet (ResNet-50)} & \textcolor{teal}{131} & 57 & 2 & 72 (32) & 5470 & 51 & 162 & 5257 (32)\tabularnewline
    {\footnotesize{}Udacity (DAVE-2)} & \textcolor{teal}{202} & 33 & 7 & 162 (24) & 2055 & 33 & 80 & 1942 (25)\tabularnewline
    {\footnotesize{}Cifar10 (ResNet56)} & \textcolor{teal}{241} & 39 & 24 & 178 (7) & 18036 & 39 & 2159 & 15838 (7)\tabularnewline
    {\footnotesize{}Cifar10 (DenseNet121)} & \textcolor{teal}{155} & 22 & 23 & 110 (5) & 1613 & 21 & 270 & 1322 (6)\tabularnewline
    \hline 
    \end{tabular}
    \par\end{centering}

    \caption{\xxx's testing time breakdown (all the numbers are in the unit of seconds). Those total testing time costs 
    comparable to the baselines are colored in green.}
\label{tab:efficiencybreakdown}
\end{table*}

\subsection{\xxx's testing time cost}\label{sec:eval:perf} 

\begin{figure}[t]
    \begin{minipage}[th]{0.4\textwidth}
        \centering
        \includegraphics[width=1.0\textwidth]{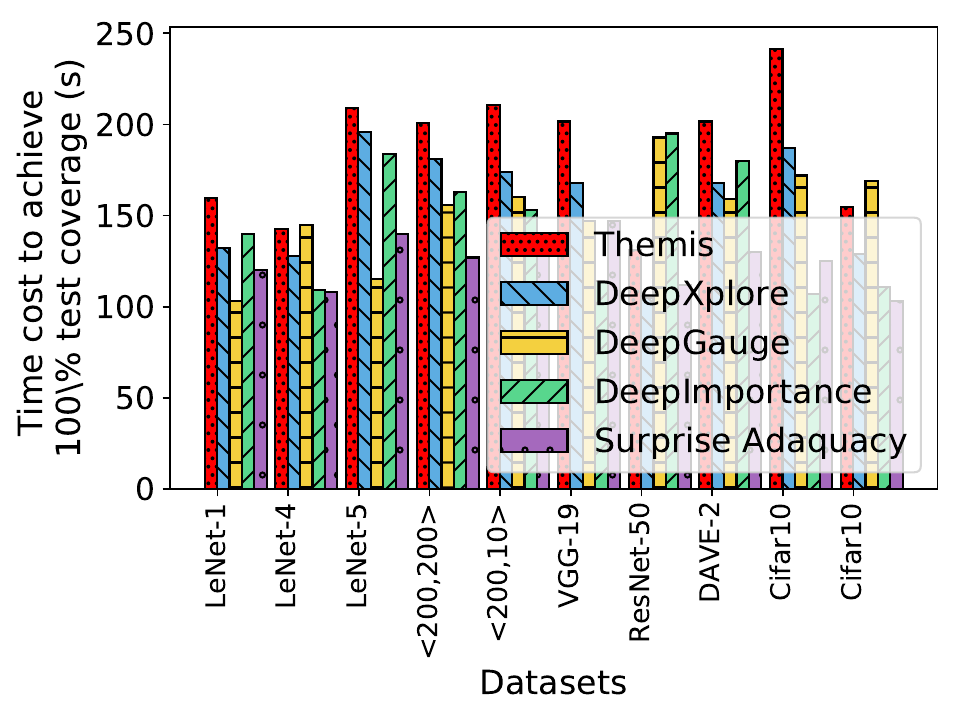}

    \end{minipage}
    \begin{minipage}[th]{0.4\textwidth}
        \centering
        \includegraphics[width=1.0\textwidth]{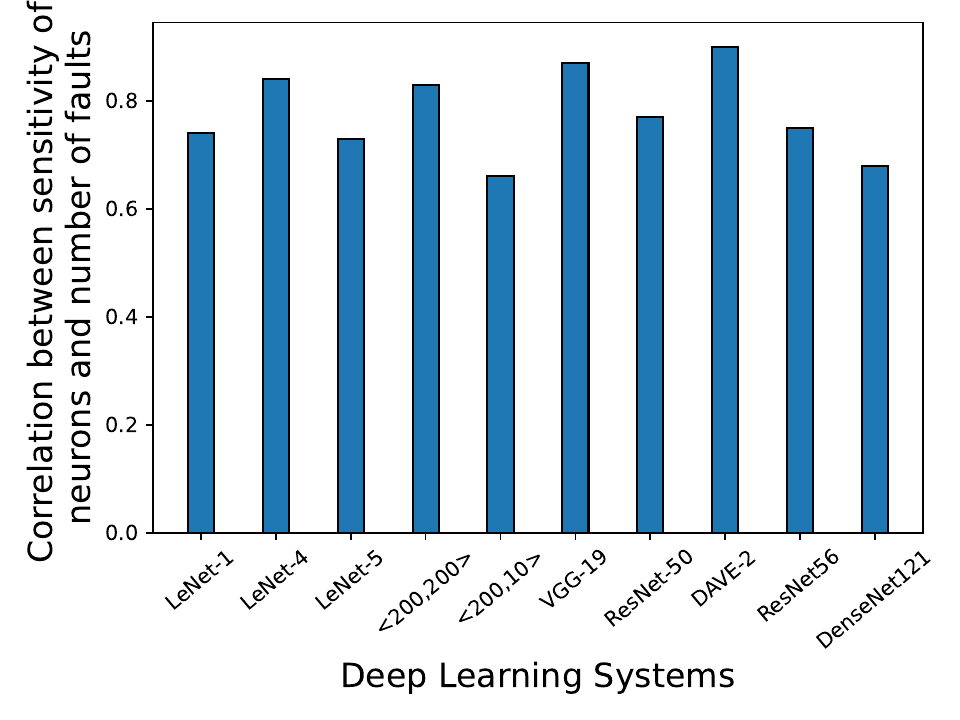}
    \end{minipage}
    \caption{Above (a): Average time taken by DLS testing 
    to complete testing (achieve 100\% test coverage).
    Bottom (b): correlation between sensitivity of neurons and the error rate.}
    \label{fig:efficiency}
\end{figure}

We then study the efficiency of \xxx. 
Figure~\ref{fig:efficiency}a shows the test time cost of
\xxx and the baselines. \xxx on average had 27.1\% more testing 
time cost than baselines. It is because \xxx was probabilistic 
and hence required more computation than deterministic 
approaches (see \ref{sec:algo:challenge}). 
Nevertheless, \xxx identified 
\numfault more faults than baselines, so the testing time 
cost is worthwhile. 

To identify the source of \xxx's time cost, we break down 
\xxx's time cost, which is comprised of three main components: 
\calculator, \coverage, and  
\fuzzer (\ref{sec:overview}). 
Table~\ref{tab:efficiencybreakdown} shows \xxx's time cost 
in \calculator, \coverage and  
\fuzzer with or without \xxx's 
\sampler. 
From the table, we can see that \xxx's performance overhead 
was mainly from \fuzzer, which iteratively performs MCMC 
(known to be time-consuming). Without \xxx's sampler component, 
\xxx's time on \fuzzer was enormous (more than one hour), because 
\xxx had to perform MCMC on all DNN's neurons. 
With \xxx's sampler, \xxx could accurately compute the 
coverage metric based on the results of one thousand 
DNN neurons (~\S\ref{sec:algo:challenge}).

\begin{figure}
    \begin{minipage}{0.4\textwidth}
        \centering
        \includegraphics[width=1.0\textwidth]{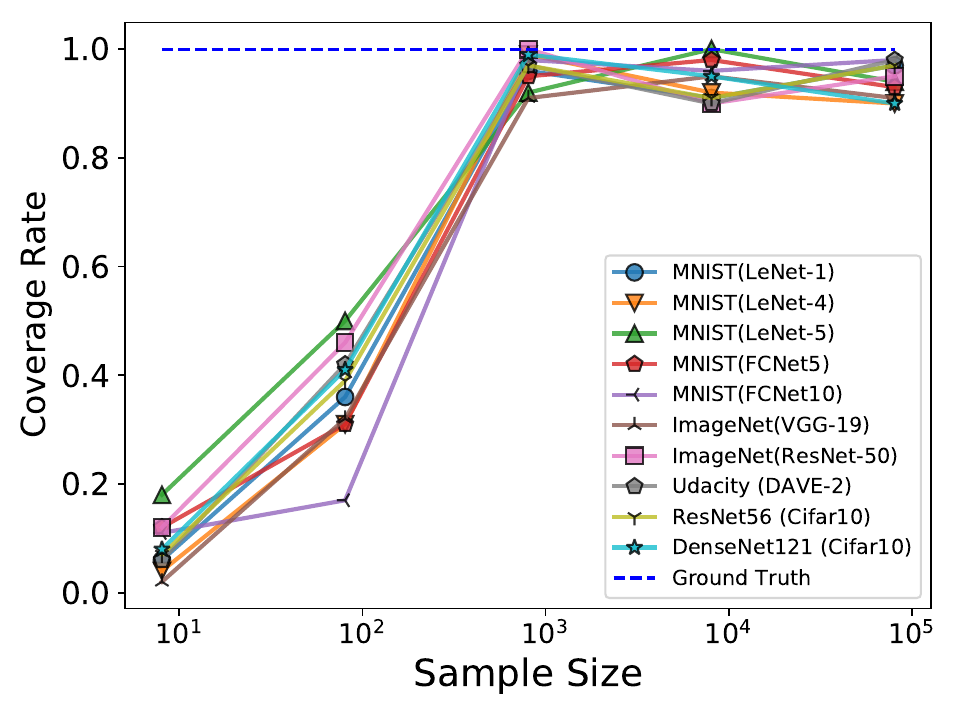}
    \end{minipage}
    \begin{minipage}{0.4\textwidth}
        \centering
        \includegraphics[width=1.0\textwidth]{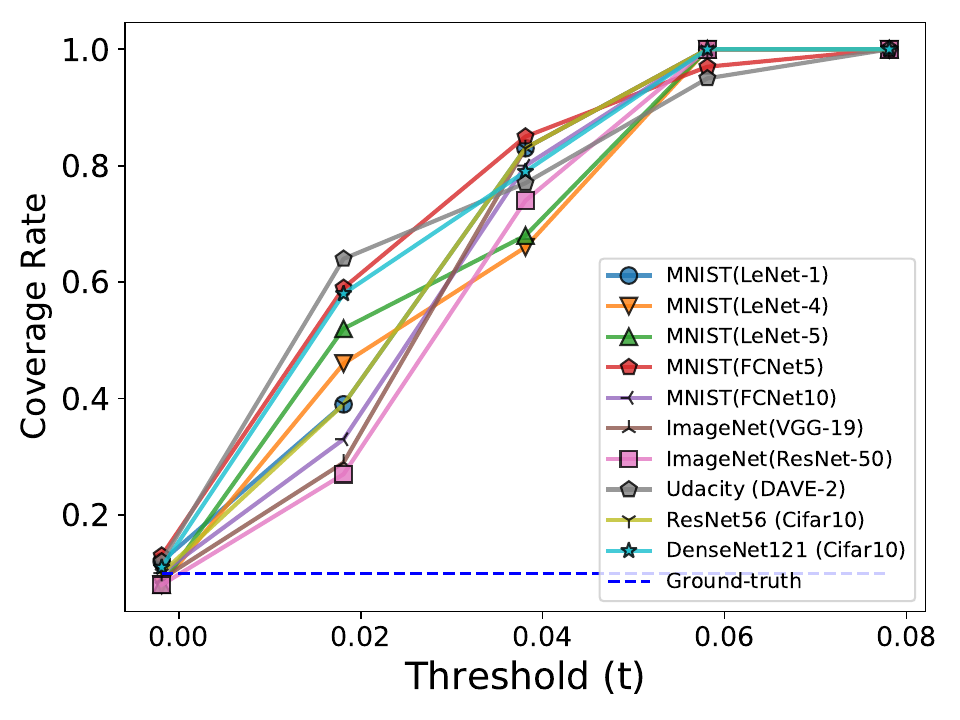}
    \end{minipage}
    \caption{Above (a): \xxx's coverage values for different sample sizes. 
    Bottom (b): \xxx's coverage values for different threshold values.}
    \label{fig:sensitivity}
\end{figure}

\subsection{Sensitivity studies on \xxx's parameters}\label{sec:eval:sens}

We study the sensitivity of \xxx's parameters (i.e., threshold $t$ and \sampler's sampler size). Figure~\ref{fig:sensitivity} 
shows \xxx's coverage variation on these values. 
Specifically, Figure~\ref{fig:sensitivity}a shows that \xxx's coverage 
was the same as the ground truth when the sample size equaled one 
thousand, which was coherent with Mann-Witney Test theory~\cite{birnbaum2020use} 
(\S\ref{sec:algo:algo}). 
We could also observe that when the sample size increased, \xxx's coverage rate was closer to the 
ground-truth. It is because statistically, with more samples, we could approximate 
the ground-truth distribution better. Nevertheless, the increase 
in precision is diminishing when the sample size grows larger. 
Hence, setting the sample size as one thousand is desirable, as justified 
theoretically (\S\ref{sec:algo:algo}) and empirically (Figure~\ref{fig:sensitivity}a).
Figure~\ref{fig:sensitivity}b shows the variation of \xxx's coverage 
against a threshold value ($t$). The figure shows that 
setting t as zero (i.e., the default value) allowed \xxx to 
achieve the ground-truth coverage value.

\subsection{Limitations and Threat to Validity}\label{sec:eval:limit}
\paragraph{limitations}
\xxx has two main limitations. First, \xxx requires extra testing time (20\%) than the baselines, because 
\xxx probabilistically tests a DNN (\S\ref{sec:eval:perf}). Nevertheless, even though \xxx has the additional testing time, \xxx still completes the testing within several minutes, which is comparable with related work~\cite{pei2017deepxplore,ma2018deepGauge,
gerasimou2020importance,kim2019guiding}. Besides, \xxx detected \numfault more faults and enhanced the accuracy of DNN on average \avgaccx times more than baselines. Hence, \xxx's additional testing time cost is worthwhile. The second limitation is that although \xxx increased a DLS's accuracy by retraining the DLS with faults detected by \xxx, \xxx does not guarantee these faults would be eliminated from the DLS after retraining. Due to the randomness nature of sampling techniques, there is a significant probability that sensitivity samples converge to a normal distribution while a tiny portion of fault-inducing data flows is not covered (\S\ref{sec:algo:algo}): the sensitivity samples can coincidentally converge to a normal distribution different from the ground-truth; hence, \xxx probabilistically (95\%, rather than deterministically 100\%) covers the fault-inducing data flows in a DLS. Indeed, how a DLS's faults can be eliminated with a guarantee is still an open challenge for all DLS testing techniques~\cite{harel2020neuron}.

\paragraph{Threats to Validity} 
The accuracy of our experimental measures and the relevance of the theoretical concepts tested are crucial for construct validity. A significant concern is whether the chosen datasets and deep learning models adequately represent the complexity of real-world scenarios. To mitigate this, we utilized a variety of well-recognized datasets (MNIST, Contagio, Drebin, ImageNet, Udacity, and CIFAR10) and deep learning systems (LeNet, VGG, ResNet, DAVE, DenseNet), each with different architectures. This diverse range ensures our findings are not limited to specific models or scenarios, thus enhancing the robustness of our construct validity.


\section{Conclusion}
This paper presents \xxx, an automated testing technique that addresses the critical challenge of fault detection in Deep Learning Systems (DLSs) used in safety-critical applications. Traditional DLS testing methods, though inspired by software testing principles, fall short in dealing with the complexities of deep learning models, particularly their sensitivity to input perturbations. \xxx overcomes these limitations by automatically exploring fault-inducing data flows, significantly reducing the reliance on manual testing efforts. This novel approach is based on the key observation that most fault-inducing data flows in DLSs are sensitive to slight changes in input. The effectiveness of \xxx is demonstrated through rigorous evaluation, showcasing its superior performance in fault detection when compared to existing DLS testing techniques. It not only achieves a higher correlation in detecting faults but also enhances the overall accuracy of DLS models. Consequently, \xxx contributes significantly to improving the reliability and efficiency of DLS applications in real-world scenarios. This advancement highlights the importance of continued innovation in AI and machine learning, ensuring these technologies meet the evolving demands of safety-critical systems in our increasingly digital world.


\section{Acknowledgements}
The work is supported in part by National Key R\&D Program of China (2022ZD0160201), HK RIF (R7030-22), HK ITF (GHP/169/20SZ), the Huawei Flagship Research Grant in 2023, HK RGC GRF (Ref.: 17208223 and 17204424), the HKU-CAS Joint Laboratory for Intelligent System Software, and the Shanghai Artificial Intelligence Laboratory, the National Research Foundation, Singapore, and the Cyber Security Agency under its National Cybersecurity R\&D Programme (NCRP25-P04-TAICeN). Any opinions, findings and conclusions or recommendations expressed in this material are those of the author(s) and do not reflect the views of National Research Foundation, Singapore and Cyber Security Agency of Singapore.

\bibliographystyle{plain}
\bibliography{acmart.bib}

\end{document}